\documentclass[12pt]{article}
\usepackage{mathrsfs}
\usepackage{amssymb}

\title{\textbf{the Half-Integer Charged Particles of the Orbifold Models}}
\author{\textbf{Cheng-Yi Sun\footnote{ddscy@163.com; cysun@mails.gucas.ac.cn}\ $^{,a}$\
and De-Hai Zhang\footnote{dhzhang@gucas.ac.cn}}\ $^{,b}$\\ \\
 {$^a$\small Institute of Modern Physics, Northwest University,}\\
     \small Xian 710069, P.R. China.\\ \\
\small$^b$Department of Physics,\\
\small Graduate School of Chinese Academy of Sciences,\\
\small Beijing 10049, P.R.China.}

\begin{document}
\maketitle
\begin{abstract}
In this paper, we consider half-integer charged particles predicted
by models of orbifold compactification of the $E_8\times E_8$
heterotic string theory. We find that it is possible for
half-integer charged particles to exist in our universe, and the
location of half-interger charged particles in a galaxy should be in
the centers of the galaxy. By qualitative analysis, we find
half-interger charged particles may be helpful in explaining the
formation of SMBH at the large redshift and solving the UHECR
puzzle.

\end{abstract}
\section{Introduction}
In the last decade, the great improvement has been made in
cosmology, which leads to the so-called ``concordant cosmological
model". The concordant cosmological model, together with the
\emph{Standard Model} of particle physics, represents our current
understanding of the nature. The later can be extended slightly by
neutrinos with nonzero masses, and the former is extended
excellently by inflation. However, there still exist lots of
questions and challenges to the concordant cosmological model and
the \emph{Standard Model}, in spite of their impressive successes.
For example, why does this particular pattern of the fundmental
interactions (the electromagnetic, weak, strong and gravitational
interactions) exist in the nature? Why is the total rank of the
electroweak-strong gauge group just four? Why are there exactly
three generations of particles?
What's the nature of the dark matter and the dark energy? What is
the mechanism that generates the baryon asymmetry? How to
incorporate the inflation with the \emph{Standard Model} properly?
Then we can conclude that the concordant cosmological model and the
\emph{Standard Model} are not the end of the story, but the tip of
the iceberg. A more fundmental theory is expected.

Several promising ideas have been put forward: Supersymmetry,
Supergravity, Superstring/M theory and Loop quantum gravity. Among
these ideas, the superstring/M theory is supposed to be the most
compelling one. This theory has some excellent properties: no UV
divergence, including gravity and gauge interaction automatically,
no free parameters before compactification, uniqueness (different
string theories are dual to each other, as a whole), etc.. However
the superstring/M theory requires a definite number of space-time,
10/11. To reconcile with our empirically 4-dimensional universe, the
extra 6/7 dimensions have to be compacted. Unfortunately, it is
argued that there may exist lots of approaches of compactification.
This even leads some authors to suggest the scenario, Landscape
\cite{landscape}. In this scenario, it is argued that there exist
plenty of vacua, at the order of $10^{500}$, as the result of
different models of compactification \cite{counting}, and the
\emph{Standard Model} can be constructed in some vacua.

Briefly, there are three classes of models on building the
\emph{Standard Model} within the framework of the superstring/M
theory: intersecting D-brane models \cite{0502005}, Calabi-Yau
models \cite{heteroticCalabiY} and orbifold models
\cite{heteroticOrbifold1,heteroticOrbifold2}. Calabi-Yau (orbifold)
models are based on the $E_8\times E_8$ heterotic superstring theory
with Calabi-Yau (orbifold) compactification. In some sense, orbifold
models can be taken as the limitation of Calabi-Yau models. Many
models on building the \emph{Standard Model} in string theory have
been suggested. Although none of the models is accepted by the
majority, we think some models are worth studying further. For
example, in Ref.\cite{0511035,kim}, some nice orbifold models have
been suggested.

In orbifold models, due to the six extra-dimensions and the lattice
group, $E_8\times E_8$, the maximum rank of gauge group is limited
to be 22, while the rank of gauge group in intersecting D-brane
models is unlimited. Then This property reduces the arbitrary of
orbifold models greatly. Usually, in the orbifold models, the
orbifold is constructed from the torus $T^6$ by identifying the
points under a discrete point group. Once the point group is given,
there are only a few possible sets of the shift vectors and the
Wilson lines from which the unequivalent breaking patterns of $E_8$
are deduced \cite{Kang-Sin Choi}. When the shift vector and the
Wilson line are given, the unbroken gauge group derived from $E_8$
is definite. Then by exhausting all the possibilities, we can select
the appropriate breaking patterns which lead the (supersymmetric)
Standard Model gauge group and matter content.

In the models of Ref.\cite{0511035,kim}, the authors obtain the 3
chiral matter generations including the right-handed neutrinos, plus
the singlet and vector-like exotic matter. Particularly, in
\cite{kim}, the authors construct a three-family flipped SU(5) model
from $Z_{12-I}$. By the Yukawa couplings, the model provides
naturally the $R$-parity, the doublet-triplet splitting, and one
pair of Higgs doublets. The model also contains some superheavy
singlets, which can be the excellent candidates for inflation
fields. So far, in the model of \cite{kim}, no serious
phenomenological problem is encountered.

But a remarkable property of this model is that it predicts
half-integer charged particles! From now on, We call half-integer
charged particle as \emph{halfon}. In fact, this is a common
property of orbifold models. However, it is well known that no
signal of the existence of fractional charged particles has been
observed. Fortunately, in the model of \cite{kim}, due to the broken
U(1) symmetries at the GUT scale, the halfons are superheavy
naturally with mass at the order of $10^{16}$Gev. Because of the
half-integer charge, the lightest of halfons, labeled by
$S^{\pm1/2}$, must be stable. Then it is possible that $S^{\pm1/2}$,
as the stable particles predicted by orbifold models, may exist in
our universe, and imprints of halfons in the universe would be
observed in the future.

However, we know the existence of fractional electric charged
particles are limited severely by observations, particularly by the
Millikan oil drop experiments. In \cite{Millikan}, it is shown that
The concentration of particles with fractional charge more than
0.16e (e being the magnitude of the electron charge) from the
nearest integer charge is less than $4.17\times 10^{-22}$ particles
per nucleon with $95\%$ confidence.

Then, in order to alleviate the limitation, generally, it is argued
that most of halfons are diluted away by inflation if the mass of
LHIC is much larger than the reheating temperature. The case is
similar to monopoles. Because density of halfons becomes so low
after inflation, we can not observe halfons on the earth. So the
contradiction between the observation and the existence of halfons
in physics is eliminated.

But, there exists another possibility that the density of halfons
may be not low. We can not observe halfons on the earth just because
halfons do not locate on the earth. Then, if halfons do not locate
on the earth, where can they locate? If halfons do exist in our
universe, are there any observable imprints? Will halfons be helpful
in solving some present puzzles in cosmology? After extensive
consideration of these questions, we find that, if we assume
appropriate halfons generated by the reheating, halfons should
condense in the center of each galaxy. Our solar system is away from
the center of the Milk Way Galaxy, so halfons can not be observed on
the earth. Further we find that, at the centers of galaxies, the
annihilation of $S^{1/2}$ and $S^{-1/2}$ can happen, and this may
explain the origin of the ultra-high energy cosmic rays (UHECRs)
above the GZK cutoff \cite{uhecr}. Even, because of the electric
charge, $S^{\pm1/2}$ may form bound states by attracting protons or
electrons. Particularly, $S^{1/2}$ and $e^-$ can form the bound
state $S^{1/2}e^-$. The state has one spectral line with the
wavelength of 4680\AA, which is slightly different from the spectral
line of Hydrogen atom with the wavelength of 4682\AA. If this
spectra line is observed, the interesting would be very great.

Below, let's show our consideration in detail. The paper is
organized as follows. In Section \ref{condensation}, by analyzing
the procession of the formation of the large scale structure, we
show that most of halfons should locate at the center of galaxies.
In Section \ref{implication}, we consider some implications of
halfons in cosmology.

\section{Condensation of Halfons}
\label{condensation}

Firstly, we assume the density of halfons generated by reheating is
appropriate. In the model of \cite{kim}, the particle content is
definite. If the inflation fields are taken as the singlets in the
model, the coupling between halfons and inflation fields can be
deduced naturally. Then halfons should be generated during reheating
by the coupling. The number density of halfons is determined by the
mass of halfon and the temperature of reheating. Then, by choosing
parameters, appropriate halfons can be generated. So we think our
assumption is reasonable. Denoting the present ratio of density of
halfons by $\Omega_S$, we expect $10^{-6}<\Omega_S<10^{-2}$. If
$\Omega_S>10^{-2}$, the big bang nucleosynthesis will be effected.
If $\Omega_S<10^{-6}$, the imprints of halfons in the universe may
be too weak to be observed.

We know the formation of the large scale structure is dominated by
the cold dark matter. Presently, a galactic halo is mainly composed
of baryonic matter and a huge cold dark matter halo. Generally, it
is supposed that the configuration of the dark matter is virialized.
The virial velocity is about at the order of $10^2km/s$. On the
other hand, the baryonic matter, due to the ability of dissipation,
forms objects of atrophysical size as individual and distinct
entities in the core of the galactic halo. Then what about halfons?


In order to answer this question, let's recall the process of the
formation of a galaxy in an ideal model. We know, the formation of
the large scale structure begins when the universe becomes the
matter-dominated. After the last scattering, the baryonic matter
falls into the well of the gravitational potential formed by the
cold dark matter. When ($\delta\rho/\rho$) becomes of order unity,
the cluster which forms the galaxy eventually, separates from the
expansion of the universe. Then the cluster begins to contract and
collapse. At the beginning of the contraction, due to the
cosmological expansion, the velocities of the cluster matter can be
taken as zero roughly. For simplicity, we assume the cluster is
roughly spherically symmetric. Then the cluster matter, including
cold dark matter, baryonic matter and halfons, begins to fall
towards the center of the cluster roughly along the radiuses under
the attraction of gravitation. If we neglect the interaction and
collision, the cluster particles would oscillate between the
antipodal points for long time. In fact, the existence of the
collision and interaction will disturb the oscillating and virialize
the cluster inevitably.
However, the process of virialization for cold dark matter, halfons
or baryonic matter is different.

For baryonic matter, due to the strong and electromagnetic
interaction, the particles of the baryonic matter collide each other
violently and frequently as falling towards the center. Then the
virialization of the baryonic matter is completed long before the
dark matter. The phase space distribution of the virialized baryonic
matter becomes roughly Maxwellian and their density varies roughly
as $r^{-2}$. Such a configuration is often referred to as an
isothermal sphere. The random distribution of the velocities of the
virialized particles would prevent the baryonic matter from
contracting further. But, due to the dissipative process--e.g.,
collisional excitation of atoms and molecules, and Compton
scattering off the CMBR, the baryonic matter can lose energy and
condense further into the core of the cluster. Finally, objects of
astrophysical size are formed as individual and distinct entities.
If the galactic halo has the angular momentum, after dissipation,
the baryonic matter will wind up in a disk-like structure
\cite{Kolb}.

For dark matter, the virialization is later than the baryonic
matter. Generally, the cold dark matter particles are supposed to be
weak-interaction massive particles (WIMPs). Due to the WIMP model,
the cold dark matter cannot be virialized by colliding with each
other or the baryonic matter. However, the galactic halo is not
spherically symmetric definitely. And the time- and pace-varying
gravitational field provides the means for WIMPs to change their
momentums and to become well mixed in phase space. Particularly, as
the condensation of the baryonic matter, Many local gravitational
centers are formed. Then the gravitational scattering of the cold
dark matter particles by the local centers accelerates the
virialization greatly. After a few dynamical times
($\tau\sim(G\rho)^{1/2}$), the virialization of the cold dark matter
is finished \cite{Kolb}. However, as the result of the WIMP model,
the virialized cold dark matter cannot lose energy to condense and
collapse further. So, now the configuration of the cold dark matter
is still a virialized halo.

For halfons, the case is different from both baryonic matter and
cold dark matter. The coupling between halfons and dark matter can
happen only by the weak neutral current, which is very weak. So
halfons almost do not collide with the cold dark matter. Since
halfon has the half-integer electric charge, it follows that halfons
can be involved in the electromagnetic interaction. So, as falling
towards the center of the cluster, halfons must collide with the
baryonic matter by the electromagnetic interaction. Due to the
superheavy mass, the effect of one collision on the motion of halfon
is unobvious. However, near the center of the cluster, the density
of the baryonic matter becomes very high. Then, near the center, the
frequent collisions between halfons and baryonic matter can
evidently reduce the kinetic energy of halfons. At the same time,
the configuration of the baryonic matter can be taken as a
isothermal sphere. So, roughly, the motion of halfon may be taken as
a damped oscillator along a radius. Then only after several
oscillations, halfons may lose most part of their kinetic energy.
The velocities of halfons become so small that they cannot move away
from the cluster center. In some sense, we can say that halfons are
virialized and become one part of the configuration of the
virialized baryonic matter at the cluster center. After then,
halfons, together with baryonic matter, condense and contract
further. Finally, halfons become one part of the individual and
distinct objects with astrophysical size near the cluster center. We
think, due to the superheavy mass, the locations of halfons should
be at the centers of these objects.

Additionally, due to electromagnetic interaction, halfons and
electrons (protons) can form some bound states, $S^{-1/2}p^+$,
$S^{1/2}e^-$, etc. These states make halfons have the ability to
undergo dissipation, as the baryonic matter, to lose energy. Even,
due to the electron/proton cloud around $S^{1/2}$/$S^{-1/2}$, the
scattering cross section of these states is much bigger than that of
halfons. Then it implies that, after having been combined with
electrons or protons, halfons collide with the baryonic matter much
more frequently. So these bound states can accelerate the
condensation of halfons.

In fact, we find that these bound states may have other remarkable
implications. For example, the ground state energy level of
$S^{1/2}e^-$ is about at the order of $10eV$, while the ground state
energy level of $S^{-1/2}p^+$ is about at the order of $10keV$. Then
the collisionally excited state of $S^{-1/2}p^+$ can emit photons
with much higher energy than that of $S^{1/2}e^-$. It implies that
$S^{-1/2}p^+$ can lose energy quicker than $S^{1/2}e^-$. Then,
finally, there may exist the segregation between the two states,
from which some new physical phenomenons may arise.

Particularly, the spectral line, $1s\rightarrow2p$, of $S^{+1/2}e^-$
is very close to the spectral line, $2s\rightarrow4p$, of $p^+e^-$.
However, due to the superheavy mass of $S^{+1/2}$, the reduced
masses of $S^{+1/2}e^-$ and $p^+e^-$ are different. So the two
spectral lines are not degenerate. The wavelength of the spectral
line of Hydrogen atom is about 4862{\AA}, while the wavelength of
the spectral line of $S^{+1/2}e^-$ is about 4860{\AA}. Then it
becomes very interesting to find whether, in our universe, there
exists the new spectral line with the wavelength 4860{\AA} or not.

Additionally, there may exist neutral states, e.g. $(S^{-1/2})^2p^+$
and $(S^{+1/2})^2e^-$. We think that it should also be interesting
to calculate the spectra of these states and then to try to find
these spectral lines in our universe.

Above, in an ideal model, we show that the location of halfons in a
galaxy should be at the center of the galaxy. For our Milky Way
galaxy, the case is more complicated and special. Our galaxy is very
huge, but the black hole at the center is small. This implies that
the formation of the Milky Way is special. The early stage of our
galaxy is the merging epoch of many dwarf galaxies, but the merging
epoch ended early. The late stage of the Milky Way is astonishingly
peaceable. In these dwarf protogalaxies, we think, halfons should
locate at the centers of the protogalaxies. Then, in the Milky Way,
most of halfons should locate in the spheroidal core of our galaxy.
Although the core of the galaxy may eject the baryonic matter during
the active epoch, few halfons may exist in the spiral arm away from
the core of the galaxy. The solar system, which may be formed by
some ejected matter, is far away from the core. So very few halfons
may exist in the solar system. Then, we think, this is the reason
that, on the earth, no signal of fractional electric charged
particles has been observed.

\section{Implications of Halfons}
\label{implication}

In the last subsection, we have given one of the implications of
halfons, the spectrum line with the wavelength 4860{\AA}. If the
spectrum line is observed, the interesting is obvious.

Additionally, We find that halfons may be helpful in solving the
UHECR puzzle. Due to the GZK cutoff, the UHECR spectrum should
dramatically steepen above $E_{GZK}\approx5\times10^{19}eV$.
However, a significant excess of events above $10^{20}eV$ has been
detected. Many proposals, including top-down models, have been
suggested to solve this puzzle (for a review, see \cite{uhecr}).
Top-down model is a generic name for all proposals in which that the
observed UHECR primaries are produced as decay products of some
superheavy particles $X$ with mass
 $m_X\gtrsim 10^{12}GeV$.
The superheavy particles may be produced by topological defects such
as cosmic strings, monopoles and hybrid defects, or be superheavy
metastable relic particles. But the both approaches have the
fine-tuning problem. For the latter, the lifetime of the superheavy
particle should be fine tuned to be in the range $10^{17}s\lesssim
\tau_X \lesssim 10^{28}s$. For topological defects, the case is even
worse, because topological defects is constrained by observation
severely.

However, if the superheavy particl is halfon, the fine-tuning
problem can be solved naturally.  Halfons are stable, and
$S^{\pm1/2}$ can annihilate into photons or $Z$ bosons,
$S^{1/2}+S^{-1/2}\rightarrow\gamma$ or $Z$. We know that it is very
difficult for the annihilation of $S^{\pm1/2}$ to happen because of
the small number density and the small annihilation cross section.
But we have shown in last subsection that most of halfons should
condense into the center of a galaxy. Then, at the centers of
galaxies, the number density of halfons may be large enough and make
it possible for $S^{\pm1/2}$ to collide with each other and then to
annihilate. The mass of $S^{\pm1/2}$ is at the order $10^{16}GeV$.
Then the annihilation may produce photons or $Z$ bosons with the
energy above $10^{24}eV$. By colliding with particles around, these
bosons can produce particles with ultra high energy (UHE) above
$10^{23}eV$ as secondaries, part of which may be UHE neutrinos or
neutralinos. These UHE neutrinos or netralinos can traverse the
extragalactic space without attenuation, thus avoiding the GZK
cutoff. Then the UHE neutrinos can collide with particles in the
galaxy, e.g. background neutrinos or neutralinos, and produce
protons with energy above $10^{20}eV$. The protons can be the UHE
primaries initiating the observed air showers and cause the UHECRs
above the GZK cutoff. Additionally, the annihilation of $S^{\pm1/2}$
happens at centers of galaxies. Then the isotropic distribution of
galaxies in the universe can explain the isotropy of UHECRs
naturally. So qualitatively, the idea works well. Of course, the
further work is needed to make sure whether halfons can solve the
UHECR puzzle quantitatively or not.

And, due to the condensation and the large mass, halfons may explain
the formation of the supermassive black hole(SMBH) at the very high
redshift. We know, the structure formation in the cold dark matter
model proceeds hierarchically, ``from the bottom up". This means
bigger structures form through tidal interaction and mergers of
smaller objects. Then the formation of SMBH at the high redshift
requires that the efficiency of the mergers should be high enough or
origin small objects should be heavy enough. Yet no one has proposed
a concrete mechanism for converting stellar mass objects into
objects 6 to 10 orders of magnitude larger or for generating origin
small objects big enough \cite{a0204486}. Halfons, because of large
mass and quick condensation, may be the natural candidate to
generate the massive seeds to form SMBH. So it is interesting for
further work to make sure whether the idea works or not.

\section{Discussion and Summary}

Above, we have shown our study on halfons. We find that halfons can
exist in out universe by condensing into the centers of galaxies.
And, we have analysed several potential implications of halfons: the
spectral line with wavelength 4860{\AA}, solving the UHECR puzzle
and explaining the formation of SMBH at the high redshift.
Superheavy halfons are the special prediction of the orbifold models
built within the framework of the superstring/M theory. If the
spectral line with wavelength 4860{\AA} is observed, it will support
strongly the orbifold models and the superstring/M theory. We know,
up to date, no observational or experimental clue in supporting the
superstring/M theory is found. This may be the first observable
signal of the superstring/M theory.

Additionally, besides the half integer charged particles, the
orbifold models also predict the superheavy singlet exotics and the
particles in the hidden sector. The singlets may be candidates as
inflatoion fields. We think that it is interesting to construct
inflation models using these singlets. The particles in the hidden
sector can deduce the broken supersymmetry and are also worth
researching.

Superheavy half integer charged particles seem to be very strange.
But we should keep our mind open. Halfon is no more strange than
axion or the dark matter. We have accepted the dark matter and
suppose the existence of axion. Why can not we assume the existence
of $S^{\pm 1/2}$?

Finally, we emphasize that, even if no signals of halfons in our
universe are observed, this does not mean that the orbifold models
should be excluded. It is possible that the number density of LHIC
is too small because of the very low reheating temperature, and then
the imprints of LHIC is too weak to be observed.

\end{document}